# How to evaluate individual researchers working in the natural and life sciences meaningfully? A proposal of methods based on percentiles of citations

Lutz Bornmann$ and Werner Marx*

$ Administrative Headquarters of the Max Planck Society, Division for Science and Innovation Studies, Research Analysis, Hofgartenstr. 8, D-80539, Munich, Tel. +49-(0)89-2108-1265, E-mail: bornmann@gv.mpg.de

* Max Planck Institute for Solid State Research, Information Retrieval Services (IVS-CPT), Heisenbergstraße 1, D-70569 Stuttgart, E-mail: w.marx@fkf.mpg.de

**Abstract**

Although bibliometrics has been a separate research field for many years, there is still no uniformity in the way bibliometric analyses are applied to individual researchers. Therefore, this study aims to set up proposals how to evaluate individual researchers working in the natural and life sciences. 2005 saw the introduction of the h index, which gives information about a researcher's productivity and the impact of his or her publications in a single number (h is the number of publications with at least h citations); however, it is not possible to cover the multidimensional complexity of research performance and to undertake inter-personal comparisons with this number. This study therefore includes recommendations for a set of indicators to be used for evaluating researchers. Our proposals relate to the selection of data on which an evaluation is based, the analysis of the data and the presentation of the results.

# 1 Introduction[1]

Researchers do science. That is why scientific success is as a rule attributed to individuals (and not institutions or research groups). As these attributions can make or break a researcher's reputation, the history of science is marked by countless disputes over the priority assigned to significant results of research (Merton, 1957). The most prestigious and best-known honour a scientist can receive today is the Nobel Prize. Every year, scientists in a number of different disciplines are awarded this prize for outstanding scientific achievement. Since governments have turned to new public management tools to ensure greater efficacy, quantitative measures of performance and benchmarking have been needed (Lamont, 2012). Prizes (not only the Nobel Prize) cannot be quantitatively analysed to provide an evaluation of the broad majority of researchers: They are rather rare events for researchers and are often awarded for achievements which lie in the distant past (Council of Canadian Academies, 2012). It has therefore become customary in the natural sciences to use bibliometric indicators to measure performance. Especially over the last few years, bibliometric assessment of individual researchers has attracted particular attention. In 2005, Hirsch (2005) presented the h index which gives information about the productivity of a scientist and the impact of his or her publications in one number (h is the number of publications with at least h citations). The h index became very popular relatively quickly (Zhang & Glänzel, 2012). However, as we show in the following, the h index is of only limited suitability for assessing a researcher's performance (Council of Canadian Academies, 2012).

Bibliometric analysis of research performance in the natural and life sciences is based on two fundamental assumptions: (1) the results of important research are published in journal articles (van Raan, 2008). That is why the number of articles which a researcher has published says something about how productive his or her research is. (2) Each new piece of research should be closely linked to current or past research (by other scientists) (Merton, 1980). These close references are marked by citations. As citations reflect the cognitive impact of the cited publication on the citing publication, the citations are considered as a measure of the impact a publication has on science.[2] It is not difficult to search the number of publications and citations listed for individual scientists in the available literature databases (Kreiman & Maunsell, 2011). Because both numbers (number of publications and citations) are linked to scientific practice[3] and the data is readily available, they have become the most important tools for evaluating individual researchers quantitatively (Garfield, 2002).

---
[1] A similar German version of this paper was published in *Zeitschrift für Evaluation*.

Today, evaluation studies go further than merely giving the number of publications and citations for a researcher; numerous bibliometric indicators are also used (Grupp & Mogee, 2004), allowing the multi-dimensional nature of scientific achievement to be captured in its complexity (Froghi et al., 2012; Haslam & Laham, 2010). Pendlebury (2009) for example suggests using eight different metrics (such as the average number of publications per year or total citation counts). Each metric has certain advantages and might compensate for the disadvantages of another (Sahel, 2011). A meaningful picture of research performance only emerges when several metrics are taken into account (Lewison, Thornicroft, Szmukler, & Tansella, 2007). However, it should be considered that many metrics chosen for a study correlate with each to a high degree (Abramo, D'Angelo, & Costa, 2010; Duffy, Jadidian, Webster, & Sandell, 2011; Hemlin, 1996) – even if results differ at a detailed level (Opthof & Wilde, 2011). Therefore, the metrics used in an evaluation study should not, as far as possible, lead to redundant results. We would like to present a selection of these metrics in this study.

---

[2] The results of studies on citing behaviour "suggest that not only the content of scientific work, but also other, in part non-scientific, factors play a role in citing behaviour. Citations can therefore be viewed as a complex, multidimensional and not a unidimensional phenomenon" (Bornmann & Daniel, 2008, p. 69). According to van Raan (2005a) "there is, however, sufficient evidence that these reference motives are not so different or 'randomly given' to such an extent that the phenomenon of citation would lose its role as a reliable measure of impact" (p. 135). A prerequisite is that the publication set of the researcher is sufficiently large.
[3] Publications and citations are linked to scientific practice to varying degrees; otherwise, we could not have *salami-slicing* or *salami style of publishing* (Bornmann & Daniel, 2007a). Scientists have been found to slice up data and interpretations into two, three, four, or more papers.

Although bibliometrics has been a separate research field for many years (Andres, 2011; de Bellis, 2009; Moed, 2005; Vinkler, 2010) there is still no uniformity in the way bibliometric analyses are applied to individual researchers (Sahel, 2011). This study aims to set up proposals how to evaluate individual researchers working in the natural and life sciences meaningfully. These proposals are particularly necessary in this area. "Evaluating individual scientific performance is an essential component of research assessment, and outcomes of such evaluations can play a key role in institutional research strategies, including funding schemes, hiring, firing, and promotions" (Sahel, 2011). Our proposals relate to the selection of data on which an evaluation of this kind is based, the analysis of the data and the presentation of the results. We have limited the study to the essential methods. This means that we only propose those (from the plethora of available options, see Vinkler, 2010) which we deem necessary and meaningful for the evaluation. Moreover, we have kept the proposals as simple as possible so that they are straightforward to use.

The following describes analyses with which to measure the productivity of a scientist and the impact of his or her research over a previous period of scientific activity. The methods proposed here are in line with the standards which we have proposed for the bibliometric analysis of research institutions (Bornmann et al., in press). To present our proposals, we use here the data for three selected researchers who work in similar areas of research but are of different ages and enjoy different levels of academic success. The data is used only to illustrate our proposals. For this reason, the researchers are designated anonymously (Person 1, Person 2 and Person 3).

# 2   Methods

**2.1  Study design**

In this section, we would like to discuss some fundamental points which should be taken into account when carrying out a study into the scientific performance of individual researchers.

1) Analysis of publications: A considerable number of publications is recommended as a basis for a statistical analysis of a single researcher. At the group level, van Raan (2000) deems 10 to 20 publications per year appropriate. According to Lehmann, Jackson, and Lautrup (2008) "it is possible to draw reliable conclusions regarding an author's citation record on the basis of approximately 50 papers" (p. 384). These recommendations for the minimum number of publications imply that an evaluated researcher should be at least at the postdoctoral level. In order to have a set that is as large as possible with which to evaluate a researcher, we recommend taking all the publications into account for the study (and not a set limited to specific publication years). This solution implies that the evaluation does not focus on the current research performance, but the performance across the whole academic career. Including all of a researcher's publications in the evaluation study obviates the need to use inference statistics to extrapolate from the selected publications (the sample) to the total number (the population) (Bornmann & Mutz, 2013).

2) <u>Citation analysis:</u> If at all possible, everything a researcher has published before the evaluation should be included in the citation analysis. However, it should also be taken into account that it is difficult to evaluate the impact of the most recent publications reliably. The most recent 1 to 2 publication years of a researcher cannot be included in the evaluation, even if methods of field normalization are used (Bornmann, 2013c; Wang, 2013). According to the Council of Canadian Academies (2012) "past research suggested that, for the natural sciences and engineering, an appropriate citation window is typically between three and five years … More recent evidence, however, has proposed that a citation window as short as two years may be appropriate in some cases … This evidence implies that citation-based indicators should be limited to assessing research published at least two years previously. Any attempt to use citation-based indicators for more recent research may result in spurious or misleading findings" (p. 68). Depending on the subject area, citations of a publication generally peak in the following two to four years before steadily decreasing in the following years. "In Biology, Biomedical research, Chemistry, Clinical medicine and Physics, the peak in citations occurs in the second year after publication, after which citations stabilize or start a decline. Citations for a second group of disciplines follow a more regular and slower-growing trend: for Earth and space science, Engineering, and especially for Mathematics, the peak of citations occurs in the last year of the time window" (Abramo, Cicero, & D'Angelo, 2011, p. 666). Therefore it is only after several years that it is possible to predict how the impact of a publication will develop.

3) <u>Self-citations:</u> In principle we are of the view that self-citations are usually an important part of the scientific communication and publication process and should therefore be taken into account in an evaluation study. "A self-citation indicates the use of own results in a new publication. Authors do this quite frequently to build upon own results, to limit the length of an article by referring to already published methodology, or simply to make own background material published in 'grey' literature visible" (Glänzel, Debackere, Thijs, & Schubert, 2006, p. 265). Only if the question of an evaluation study explicitly means to what extent a scientist has influenced other scientists' work, self-citations should be obviously ignored.

In every evaluation study, however, it should be checked whether a researcher cites him or herself excessively. A large study examined the proportion of author self-citations in Norway (from 1981 to 1996): "More than 45,000 publications have been analysed. Using a three-year citation window we find that 36% of all citations represent author self-citations" (Aksnes, 2003, p. 235). Our experience in practical evaluation in the natural and life sciences has shown that the percentage of self-citations is 10-20%. Given the information in the Norwegian study and similar data in other publications (Andres, 2011, pp. 60-61) and our experience in compiling bibliometric reports for individual researchers, we think that a figure that does not exceed approximately 30% is a reasonable level of self-citation (van Raan, 2005b).

## 2.2 Describing the researcher

If possible, a study evaluating an individual researcher should include information about his or her career so that the bibliometric results can be interpreted against this background (Cronin & Meho, 2007; Sugimoto & Cronin, 2012). This information includes, for example, the institutions where a researcher has already worked or is currently working. If the researcher has a web site, the URL should be given in the evaluation report. The following provides some help regarding other bibliographical information: "For each scientist, we gathered employment and basic demographic data from CVs, sometimes complemented by Who's Who profiles or faculty web pages. We record the following information: degrees (MD, PhD, or MD/PhD); year of graduation; mentors during graduate school or post-doctoral fellowship; gender; and department(s)" (Azoulay, Graff Zivin, & Manso, 2009, p. 14). There are similar descriptions in other studies (Duffy, et al., 2011).

We do not supply any biographical information for the three researchers who have been included as examples in this study in order to preserve their identity.

**2.3   Description of the database**

As many names in the literature databases (such as Smith, A.) cannot be assigned completely unambiguously to a certain person, compiling the publication set so that it is completely reliable represents a major challenge for single researcher evaluation studies. "In bibliometrics, name ambiguity represents a considerable source of error and can affect the quality and validity of the results" (D'Angelo, Giuffrida, & Abramo, 2011, p. 258). It is estimated that at least 10% of authors share their name with one or more other authors (D'Angelo, et al., 2011; Strotmann & Zhao, 2012). It would be very helpful for the evaluation process if each researcher had a unique identification number through which every publication could be accessed. Initiatives of this nature already exist (see for example http://www.researcherid.com/), but they have not yet had reliable and definite results for all researchers (Kreiman & Maunsell, 2011).

The best approach to recording publications accurately would therefore be to use personal publication lists. However, in many cases this is not possible for practical reasons. We therefore recommend that the publications be searched in the databases and – in order to avoid errors and omissions – that the searched publication sets be cross-checked against the publications proven to be from the researchers in question (on his/her institutional homepage). This cross-check should at least cover whether the number of the publications searched in the databases matches the number given by the researcher. Where there are differences, the search strategy in the database should be optimised or the information provided by the researcher verified (Bornmann, et al., in press). In general, researchers should be best possibly involved in the process of compiling their publication set. There may be exceptional circumstances where this may not be the case, but one would usually expect that the individual scientist can provide the personal publication list.

The databases used as a rule in evaluative bibliometrics are those supplied by Thomson Reuters (Web of Science, WoS) and Elsevier (Scopus) (Council of Canadian Academies, 2012). In some disciplines it is advisable to work with specialist databases (as well). Some of these now give the citation counts for publications as well (for example, Chemical Abstracts in Chemistry). However, we do not advise using Google Scholar (GS) as a basis for bibliometric analysis. Several studies have pointed out that GS has numerous deficiencies for research evaluation (Bornmann et al., 2009; García-Pérez, 2010; Jacso, 2009, 2010). Besides a number of bibliometric centres (e.g., the Centre for Science and Technology Studies, CWTS, in Leiden, the Netherlands), Thomson Reuters is one supplier of relative citation rates which are time and subject-normalised and which can be used for bibliometric-based evaluation of research. The relative (that is, time and subject-normalised) citation rates can be obtained from the National Citation Report and InCites. Both databases are based on the WoS. Independent of the source of the time and subject-normalised data (Thomson Reuters or bibliometric centre), the quality of the data should be checked accordingly: errors in normalization can strongly bias the analysis at the level of the individual. To test the reliability of the data, the empirical results based on two different data sources can be compared.

The number of publications in the WoS core journals (currently around 10,000 fully recorded journals) has become the standard measure for the quantification of scientific productivity in the natural and life sciences. In WoS, Thomson Reuters offers various citation indexes (such as the Science Citation Index, SCI, and the Social Science Citation Index, SSCI), the availability of which is subject to a licence and which should therefore be documented in every study. Not only outside of the core natural and life sciences subjects (particularly in the area of computer science and engineering science and technology) but also inside of these subjects the coverage of the publications by the databases is not 100%. Moed (2005, p. 126) presents percentages for the overall coverage of the natural and life sciences by the SCI, which range from 92% (molecular biology and biochemistry) to 53% (mathematics) (see also Korevaar & Moed, 1996; Moed & Hesselink, 1996). This means that in the best case, an average of 8% is missing, in the worst case an average of 47%. A similar range was found by Butler and Visser (2006, p. 329) for Australian university publications. It should be further considered that the coverage of publications by WoS is shrinking (Larsen & von Ins, 2009).

**2.4 Software**

We used the statistics program Stata (Bornmann & Ozimek, 2012; StataCorp., 2011) to analyse the data for this study. Other programs (such as SPSS or R) can also be used for such analyses. The results are presented in line with the American Psychological Association (2010) guidelines, the standard in empirical social sciences.

# 3 Results

A summary of the productivity and citation impact results for the three scientists is shown in Table 1. The detailed results for each indicator are presented in additional tables and illustrations. The productivity indicators (upper part of Table 1) aim to present a differentiated picture of the publication output, in particular with regard to the document types and the author succession. The impact indicators (lower part of Table 1) include three types: (1) the base data (total citations, proportion of self-citations, citations per publication), (2) the widely used h index and one of its variant, the m quotient, and (3) the normalized impact indicators (average percentile, top 10% based excellence indicators). These indicator types provide quite different kinds of information concerning scientific performance: The total number of citations and the number of citations per publication are hardly more than the raw data for the calculation of advanced impact indicators. The proportion of self-citations shows whether a critical upper limit is exceeded by a researcher.

The h index (and also the m quotient) has to be included because of its proliferation within the scientific community. The advanced indicators, i.e. the normalized impact indicators, provide a time period- and field-independent measurement of research performance, which focus on a researcher's complete set of publications or alternatively on the amount of excellent publications. Since the various impact indicators measure research performance differently, there numbers may contradict each other. We suggest to give the advanced indicators, the (average) percentiles and the top 10% based excellence indicators, the highest weight when comparing the scientific performance of single researchers. Only these indicators facilitate a fair evaluation of performance. Even if advanced indicators are used, it is important that the results are interpreted against the backdrop of each researcher's curriculum vitae and the specific needs of the evaluation process.

### 3.1 Productivity

*Publications*

Figure 1 shows for each researcher the number of publications by document type (also see Table 1). Note that the Thomson Reuters classification of publications by document types follows their own criteria and frequently is not in line with the classification in the journals (Meho & Spurgin, 2005). When published, original research results are usually classified by database producers as "Articles" and long literature overviews as "Reviews" (Moed, van Leeuwen, & Reedijk, 1996). As Figure 1 shows, publications with the document type "Article", dominate for all three researchers. "Proceedings papers" also play an important part, particularly for researcher 3, but also for the other two researchers. Researcher 1 has published significantly more documents of all types (n=190) than Researcher 2 (n=76) and 3 (n=95).

While Researcher 1 published 7.9% (n=15) of his publications as first author (he is the sole author in none of them), this figure is 22.4% of Researcher 2's publications (n=17) (he is the sole author in 5 of them) and for Researcher 3 it is 40% (n=38) of the publications (in 12 of which he is the sole author) (see Table 1). We recommend that this information about authorship is taken into account when comparing the productivity of researchers (Sugimoto & Cronin, 2012; Zhang & Glänzel, 2012). A publication written without co-authors generally requires more work than one with co-authors (Kreiman & Maunsell, 2011). Furthermore, publications in which the scientists are first authors can be considered more significant in most disciplines, as the first authors frequently do most of the research (de Moya-Anegón, Guerrero-Bote, Bornmann, & Moed, in press).

In addition to authorship and document types, the distribution of publications over the years is also an interesting factor in researcher evaluation. Are the publications distributed evenly or unevenly? Does productivity increase or decrease; that is, is there a noticeable trend over the years? When did the academic career start? As a rule, this is considered equivalent to the appearance of the first publication (Kreiman & Maunsell, 2011). As Figure 2 shows for the researchers investigated in this study, publishing history can vary widely (also see Table 1): While Researchers 1 and 3 published for the first time as early as the beginning of the 1990s, Researcher 2 started much later, in 2001. Researcher 1 achieved the highest levels of productivity approximately 10 years after the beginning of his/her academic career and since then has published around 5 times a year. Since the start of his or her career, Researcher 2's publications have demonstrated a rising trend which stabilised at 14 per year between 2009 and 2011. Researcher 3 published at a consistently low level from the beginning of the 1980s, peaking at 10 publications in 1997 over many years of publishing activity. As the results of summarizing analyses show in Table 1, Researcher 1 has 5.9 publications per year (arithmetic average); this figure is 6.9 for Researcher 2 and 3.2 for Researcher 3.

*Journals*

According to Pinski and Narin (1976) the analysis of researcher productivity does not take account of the importance of their publications: "The total number of publications of an individual, school or country is a measure of total activity only; no inferences concerning importance may be drawn" (p. 298). There should therefore be additional analysis to reveal the significance of the publications. As well as the citation analysis for each publication, which is shown in the next section, we recommend listing the journals in which a researcher has published. The Normalized Journal Position (NJP) should also be given so that the importance of the journals in their subject area can be determined. We recommend using this indicator rather than the Journal Impact Factor (JIF) as it is not possible to compare the JIFs of journals in different fields with each other (Bornmann, Marx, Gasparyan, & Kitas, 2012; Marx & Bornmann, 2012; Pendlebury, 2009). The NJP is a gauge of the ranking of a journal in a subject category (sorted by JIF) to which the journal is assigned by Thomson Reuters in the Journal Citation Reports (JCR) (if a journal belongs to more than one category, an average ranking is calculated). "Unlike the IF med [Median JIF of publications], it [NJP] allows for inter-field comparisons as it is a field-normalized indicator" (Costas, van Leeuwen, & Bordons, 2010, p. 1567). The lower the NJP for a journal, the higher its impact in the field.[4]

It is not possible to include all the publications from the three researchers in the calculation of the NJP. Only those articles can be taken into account which have been published in journals currently analysed by Thomson Reuters for the JCR (and for which a JIF is calculated). 169 publications by Researcher 1 (89%), 63 publications by Researcher 2 (83%) and 85 publications by Researcher 3 (90%) are included in the analysis. JIFs from the JCR Science Edition 2011 were used to calculate the NJP. For example, Thomson Reuters assigns the journal *Chemistry of Materials* to the subject categories "Chemistry, Physical" and "Materials Science, Multidisciplinary." In "Chemistry, Physical" the journal ranks 14 in a total of 134 journals (sorted in decreasing order by the JIF for 2011)(14/134=0.105) and in "Materials Science, Multidisciplinary" ranks 13 in a total of 231 (13/231=0.056). The NJP for this journal is 0.08 ((0.105+0.056)/2).

---

[4] As an alternative to the NJP, other methods for normalizing the JIF could be used. An overview of these methods can be found in Vinkler (2010, pp. 186-189). However, it is an advantage of the NJP compared to the other methods that the necessary data are readily available in the WoS.
It might be a disadvantage of all normalizing methods that they are based on journal sets to delineate different fields. It is well known that these categories can be quite imprecise – especially in case of multi-disciplinary journals and highly specialized fields of research (Bornmann, Mutz, Neuhaus, & Daniel, 2008). Thus, if a publication list contains publications from these journals and/or the evaluated scientist is active in a highly specialized field, the use of journal metrics based on journal sets may be a problem.

The results of the analysis of journals for the three scientists are shown in Table 2. Researcher 1 has the most publications (n=72) in *Journal 14* with an NJP of 0.19; for Researcher 2 the most publications (n=9) are in *Journal 3* with an NJP of 0.03 and for Researcher 3 (n=34) in *Journal 21* with an NJP of 0.31. The best NJP for all the scientists is 0.01 for *Journal 1*. Taking an average over all the journals, the NJP is better for Researcher 2 at 0.19 than for Researcher 3 (NJP=0.29) and Researcher 1 (NJP=0.36). The average impact for the journals in which Researcher 2 has published is thus higher than for Researchers 3 and 1.

### 3.2  Impact

*Citations*

Citations measure an aspect of scientific quality – the impact of publications (van Raan, 1996). Martin and Irvine (1983) distinguish between this aspect ("the 'impact' of a publication describes its *actual* influence on surrounding research activities at a given time," p. 70) and 'importance' ("the influence on the advance of scientific knowledge," p. 70) and 'quality' ("how well the research has been done," p. 70). They consider the impact as the most important indicator of the significance of a publication on scientific activities. Cole (1992) sees citations as a valid indicator of quality, as they correlate with other quality indicators: "Extensive past research indicates that citations are a valid indicator of the subjective assessment of quality by the scientific community. The number of citations is highly correlated with all other measures of quality that sociologists of science employ. As long as we keep in mind that research of high quality is being defined as research that other scientists find useful in their current work, citations provide a satisfactory indicator" (p. 221, see also Bornmann, 2011; Smith & Eysenck, 2002). Other benefits of citations for measuring quality (using the impact) are (Marx, 2011): "it is valid, relatively objective, and, with existing databases and search tools, straightforward to compute" (Nosek et al., 2010, p. 1292).

While we have taken account of all the document types in the analyses of productivity (see above), it is recommended that only substantial works of research are included in citation analyses: "The standard practice is to use journal items that have been coded as regular discovery accounts [articles], brief communications (notes), and review articles – in other words, those types of papers that contain substantive scientific information. Traditionally left to the side are meeting abstracts (generally not much cited), letters to the editor (often expressions of opinion), and correction notices" (Pendlebury, 2008). Following this recommendation, the results presented in the following encompass only "Articles", "Notes", "Proceedings Papers" and "Reviews" by the three researchers. In total, there are 15,192 citations for Researcher 1, 3,796 for Researcher 2 and 7,828 for Researcher 3 (see Table 1).[5] While the proportion of self-citations among these citations for Researcher 1 is 3-4%, this value is approximately 6% for Researchers 2 and 3. On average, Researchers 1 (M=83) and 3 (M=89) have had significantly more citations per publication than Researcher 2 (M=52)

*Percentiles*

Numerous studies in bibliometrics have shown that citation counts are time and field-dependent. We can therefore expect a varying number of citations for publications in different fields and years. "This is due to a number of factors: (i) different numbers of journals indexed for the fields in the main bibliometric databases, such as Web of Science or Scopus; (ii) different citation practices among fields and last, but not least (iii) different production functions across fields" (Abramo, et al., 2011, p. 661). According to (Schubert and Braun (1993), 1996)), normalisation should therefore be used with citation analyses. Current research into bibliometrics indicates that there are good arguments in favour of percentiles for normalising citations of individual publications in terms of the subject area and the publication year (Bornmann, Mutz, Marx, Schier, & Daniel, 2011; Leydesdorff, Bornmann, Mutz, & Opthof, 2011; Waltman, et al., 2012b). "First, it [percentile ranking] provides normalization across time such that papers from different years can be directly compared. This result is particularly important for recent papers, because they have typically not had enough time after publication to accumulate large numbers of citations. Second, given the skewed nature of citation count distributions, it keeps a few highly cited papers from dominating citation statistics" (Boyack, 2004, p. 5194; Ruiz-Castillo, 2012). According to analyses by Albarrán and Ruiz-Castillo (2011), around 70% of the publications in a set receive fewer citations than average and 9% of the publications can be designated as highly-cited.

---

[5] Citations are a probabilistic process and therefore the number of citations to the publications of the researchers may vary for all sorts of reasons that have nothing to do with cognitive impact (Bornmann & Daniel, 2008). In addition, the measurement of citations does inevitably entail measurement errors. Hence, statistical estimations of the possible error involved – like confidence intervals (Cumming, 2012) or stability intervals (Waltman et al., 2012b) – around the values of citation indicators could be calculated and added.

The percentile provides information about the impact the publication in question has had compared to other publications (in the same subject area and publication year). Using the distribution of citation frequencies (sorted in descending order) all the publications from the same subject area and the same year of publication as the publication in question are broken down into 100 percentile ranks. The maximum value is 100 which denotes that a publication has received 0 citations (based on the InCites percentile definition). Accordingly, the lower the percentile rank for a publication turns out to be, the more citations it has received among the publications in the same subject area and publication year. The percentile for the publication in question is determined using the distribution of the percentile ranks over all publications. For example, a value of 10 means that the publication in question is among the 10% most cited publications; the other 90% of the publications have achieved less impact. A value of 50 represents the median and therefore an average impact compared to the other publications. Normalising citations with percentiles allows the impact of publications from different subject areas and publication years to be compared with each other.

Figure 3 shows the distribution of percentiles for the publications which the three researchers have published over the years. Beam plots (Doane & Tracy, 2000) have been used for illustration (see Bornmann & Marx, in press). They make it possible to present the distribution of percentiles in a publication year combined with the median from these percentiles. It is an advantage of beam plots that they allow two perspectives of evaluations: an overview on the whole career of a person and the focus on specific time periods during the career (e.g., on current research activities) for standalone assessment as well as for comparisons between scientists. While in Figure 3 the individual percentiles for the publications are shown using grey rhombi, the median over a year is displayed with the aid of a red triangle. Furthermore, for each person a red dashed line shows the median of the percentiles for all the years and a grey line marks the value 50. As described above, a value of 50 designates the average impact of a publication in a subject area or publication year. The percentiles for 2011 are only included in order to show all the publication years; as the citation window for these publications is as a rule too short to accumulate citations, the percentile in many cases is 100 (see Researcher 2 in the figure, for example).

As the analyses for the three researchers in Figure 3 show, they achieved a very substantial impact with their publications on average (median). While Researchers 2 and 3 have an average percentile of 6.2 and 8.3, for Researcher 1 this figure is 15.9 (see Table 1). Apart from 2005 and 2006, Researcher 2 has had average percentiles around a value of 10 since he or she began publishing. This makes these publications among the 10% most cited publications in their subject area and publication year. Researcher 3 exhibits a similarly excellent performance over the last twenty years of his or her publishing activity.

## 3.3 The combination of the number of publications and their impact in one number

In 2005, the h index was proposed as an indicator with which to measure the performance of individual researchers as follows: "A scientist has index $h$ if $h$ of his or her $N_p$ papers have at least $h$ citations each and the other $(N_p - h)$ papers have $\leq h$ citations each" (Hirsch, 2005, p. 16569). Although before 2005 the performance of researchers was usually measured with separate indicators for productivity and impact, the h index combines both of these into one number. The h index was adopted relatively quickly by science insiders and non-academics and became the subject of discussion (Bornmann & Daniel, 2007b, 2009). By the end of 2011, Hirsch's (2005) publication had been cited almost 1000 times. The h index is now offered as an indicator in many literature databases, such as WoS and Scopus. In so far as bibliometrics has studied the h index, it is concerned primarily with the advantages and disadvantages of the h index (Alonso, Cabrerizo, Herrera-Viedma, & Herrera, 2009; Egghe, 2010; Norris & Oppenheim, 2010; Panaretos & Malesios, 2009; Thompson, Callen, & Nahata, 2009; Zhang & Glänzel, 2012). On the one hand, it is seen as an advantage that the h index is easy to calculate, but on the other, a disadvantage that it is normalised neither for age nor for field. It is not possible to compare the h index of researchers from different fields and of different (academic) ages with each other. Against the background of the disadvantages of the h index, almost 40 variations on the h index such as Egghe's (2006) g index have been proposed (Bornmann, Mutz, Hug, & Daniel, 2011). However, none of the variations have so far prevailed successfully over (or besides) the h index.

As Table 1 shows, Researcher 1 has a significantly higher h index (h=54) than Researcher 2 (h=27) and Researcher 3 (h=38). As the h index depends very much on the productivity and/or the (academic) age of a researcher (Bornmann, Mutz, & Daniel, 2008), we followed the recommendation by Hirsch (2005) and have normalised the h index for age, by dividing it by the number of years since the appearance of the first publication. Hirsch (2005) calls this quotient the m quotient. For the three researchers (see Table 1) it reveals a clear advantage in the performance of Researcher 2 (m=2.5) compared to Researcher 1 (m=1.7) and Researcher 3 (m=1.2). Even though the h index is age-normalised to give the m quotient, the second step, normalisation for field is missing. (Bornmann (2013a), 2013b)) therefore suggests an alternative to the h index: specifying the number of publications for a researcher which belong to the 10% of the most-cited publications in their field and publication year ($P_{top\ 10\%}$). This indicator is based on the percentile approach, in that it counts those publications with a percentile less than or equal to 10 (see above). The indicator is one of the "success indicators" in bibliometrics which count successful publications and take time and field-normalisation into account (Franceschini, Galetto, Maisano, & Mastrogiacomo, 2012; Kosmulski, 2011, 2012).

As well as field-normalisation, $P_{top\ 10\%}$ offers another advantage in that it does not use an arbitrary threshold to determine publications in a set with high citation impact. A number of publications (Waltman & van Eck, 2012) have already pointed out the disadvantage of this arbitrariness with the h index. "For instance, the h-index could equally well have been defined as follows: A scientist has an h-index of h if h of his publications each have at least 2h citations and his remaining publications each have fewer than 2(h+1) citations. Or the following definition could have been proposed: A scientist has an h-index of h if h of his publications each have at least h/2 citations and his remaining publications each have fewer than (h+1)/2 citations" (Waltman & van Eck, 2012, p. 408). According to Kreiman and Maunsell (2011), a threshold should be defined as follows: "This threshold would have to be defined empirically and may itself be field-dependent. This may help encourage scientists to devote more time thinking about and creating excellence rather than wasting everyone's time with publications that few consider valuable." A standard in bibliometrics is used to select highly cited publications for $P_{top\ 10\%}$: Publications which are among the 10% most cited publications in their subject area are as a rule called highly cited or excellent (Bornmann, de Moya Anegón, & Leydesdorff, 2012; Sahel, 2011; Tijssen & van Leeuwen, 2006; Tijssen, Visser, & van Leeuwen, 2002; Waltman, et al., 2012b).

As the analyses of the $P_{top\ 10\%}$ for the three researchers in Table 1 show, Researcher 1 has many more excellent publications (70) than Researchers 2 ($P_{top\ 10\%}$ = 31) and 3 ($P_{top\ 10\%}$ = 48). To compare the number of $P_{top\ 10\%}$ with an expected value, it is possible to calculate the proportion of $P_{top\ 10\%}$ in a researcher's publication set ($PP_{top\ 10\%}$). A comparison with an expected value is not possible with the h index. The expected value of $PP_{top\ 10\%}$ is 10%. If one were to select sample publications (percentiles) at random from a database, such as InCites, it could be expected that 10% of the publications would belong to the 10% of the most cited publications in their subject area and publication year (Bornmann, de Moya Anegón, et al., 2012). $PP_{top\ 10\%}$ is seen as the most important indicator in the Leiden Ranking by the Centre for Science and Technology Studies (Leiden University, The Netherlands): "We … regard the $PP_{top\ 10\%}$ indicator as the most important impact indicator in the Leiden Ranking" (Waltman et al., 2012a, p. 10). As Table 1 shows, all three researchers have considerably more highly-cited publications than might be expected. For Researchers 2 and 3, even more than half of the publications are in $P_{top\ 10\%}$.

In the same way as Hirsch (2005) proposed the m quotient for the h index, we would like to propose using the number of years as an active researcher ($P_{top\ 10\%}$ quotient) to normalise $P_{top\ 10\%}$ for age. Indicators for individual researchers should in general be normalised for age. It is possible to explain the cumulative impact of publications by a researcher to a great extent by the years since completion of his or her doctoral studies: "Years since PhD accounted for 43% of the variance in log(total citations), 48% of the variance in log(*h*), 36% of the variance in log(*e*), and 54% of the variance in log($h_m$) [*e* and $h_m$ are variants of the h index]" (Nosek, et al., 2010, p. 1287). In taking into account the number of years as an active researcher, the $P_{top\ 10\%}$ quotient is therefore normalised not just in terms of the publication year and the field of the individual publications (see above), but also in terms of the age of the researcher. The results with this indicator are shown in Table 1. With a value of 2.8, Researcher 2 published around twice as many $P_{top\ 10\%}$ as Researcher 3 ($P_{top\ 10\%}$ quotient=1.6). The $P_{top\ 10\%}$ quotient for Researcher 1 is 2.2.

## 4 Discussion

An evaluation report for one or more researchers should conclude with a short summary of the most important results. Although with 3 publications per year Researcher 3 is the least productive of the three, (the other two researchers have published around 6 times a year), he or she has produced by far the most publications as first author or single author (38 and 12 respectively). The average impact of the journals in which Researcher 2 has published is higher than that of Researchers 3 and 1. Researcher 1's publications have been cited most (n=15,192). Researcher 2 does very well particularly on the age-normalised indicators: His or her m quotient (2.5) and $P_{top\ 10\%}$ quotient (2.8) are significantly higher than those of the other two researchers. At 57.8%, Researcher 3 has the highest proportion of excellent publications ($PP_{top\ 10\%}$) in the set.

In this study, we have endeavoured to present a set of different bibliometrical methods with which to evaluate a single researcher. This set is flexible and can be adapted to the application in question. The methods and indicators presented here need not be used in every instance. For example, with the indicators based on $P_{top\ 10\%}$ which we have presented for showing publication impact the focus is on excellence: the ability of researchers to (a) publish in excellent journals (that is, journals which achieve on average a high impact in their discipline) and (b) produce publications which are cited very frequently compared to other publications in the same field (Tijssen & van Leeuwen, 2006). The focus on excellence is in line with a general trend in science policy: "Many countries are moving towards research policies that emphasize excellence; consequently, they develop evaluation systems to identify universities, research groups, and researchers that can be said to be 'excellent'" (Danell, 2011,

p. 50). Moreover, a trans-disciplinary bibliometric study could show that scientific progress is based primarily on highly-cited publications (Bornmann, de Moya-Anegón, & Leydesdorff, 2010). If, however, an evaluation of a single scientist does not focus on excellence, the impact analyses could be restricted to the presentation of beam plots.

Percentiles are used to normalise the impact of individual publications for time and subject area. It is this normalisation which makes it possible to make meaningful statements about the impact of publications. However, the normalisations are carried out on the level of the individual publications and are limited to the impact of individual publications. In order to make it possible to make evaluative statements about the productivity and impact of a person, it would be desirable to have available benchmarks at the individual level. Kreiman and Maunsell (2011) have already said as much (Garfield, 1979): "When comparing different post-doctoral candidates for a junior faculty position, it would be desirable to know the distribution of values for a given index across a large population of individuals in the same field and at the same career stage so that differences among candidates can be evaluated in the context of this distribution. Routinely providing a confidence interval with an index of performance will reveal when individuals are statistically indistinguishable and reduce the chances of misuse" (p. 249). While in many disciplines there are no such comparison values, they have already been introduced in the fields of logistics and medical informatics to evaluate the productivity and impact of researchers (Coleman, Bolumole, & Frankel, 2012; El Emam, Arbuckle, Jonker, & Anderson, 2012).

When a researcher is evaluated, the bibliometric analyses should be supplemented with the analysis of other indicators. "It also strongly recommended that additional criteria be taken into consideration when assessing individual research performance. These criteria include teaching, mentoring, participation in collective tasks, and collaboration-building, in addition to quantitative parameters that are not measured by bibliometrics, such as number of patents, speaker invitations, international contracts, distinctions, and technology transfers" (Sahel, 2011). Bibliometrics needs to be enhanced as appropriate (or replaced by other indicators) particularly in disciplines which cannot be included among the natural and life sciences. "For the humanities and social sciences (philosophy, history, law, sociology, psychology, languages, political sciences, and art) and for mathematics, the existing databases do not cover these fields sufficiently. As a consequence, these fields are not able to properly use bibliometrics" (Sahel, 2011).

An expert in bibliometrics (familiar with research evaluation) should decide in every case how a researcher is evaluated bibliometrically. A qualified expert has published in this field and should follow standard procedures. Bibliometrics is now a field in its own right with its own specialist journals and regular conferences. "Calculations should not be left to non-specialists (such as administrators that could use the rapidly accessible data in a biased way) because the number of potential errors in judgment and assessment is too large. Frequent errors to be avoided include the homonyms, variations in the use of name initials, and the use of incomplete databases" (Sahel, 2011). Only experts in bibliometrics can take account of the diverse problems and difficulties which can arise in a bibliometric analysis (Retzer & Jurasinski, 2009). In principle, the evaluation of a researcher should be carried out as part of an "informed peer review" (Abramo & D'Angelo, 2011; Taylor, 2011). This involves referees from the same discipline as the researcher being evaluated. We see it as the task of the experts in bibliometrics to give guidance to the referees by indicating how to interpret the different indicators and the results of the bibliometric evaluation. The referees produce a review on the basis of (i) their own assessment of the researcher and (ii) a bibliometric analysis (undertaken in advance by an expert in bibliometrics).

Quantitative (bibliometric) methods for measuring the productivity and impact of research performance are particularly at risk from the incorrect interpretation of data. This is because research into the underlying data is usually separate from its interpretation and application for the purposes of evaluating research. It should not be forgotten that much bibliometric data is politically critical and associated with strong interests (in particular reputation and money). Bibliometric indicators have become such a powerful tool within the context of science policy that consideration must be given to their potential for misleading and destructive use. Their potency requires a code of professional ethics to govern their application (Weingart, 2005). Primarily this means applying the best and fairest approach available in the current bibliometric community (that is, the most appropriate indicators and not the simplest and cheapest) and also that the limitations of the method and potential distortions are pointed out (Marx & Bornmann, in press).

Scientists, who should be used to handling bibliometric data as end users, should be able to understand the limitations of the data and the risks that can result and it must be possible for them to call them to account. However this is often not the case: when money and reputation are at stake, scientists are also only human and forget the rules of good scientific practice. Bibliometric data is likely to be misinterpreted if this can benefit their positive image or completely ignored if it does not provide confirmation of scientists' perception of themselves. It might also be used as ammunition against competitors if it seems appropriate for this purpose. The danger of partiality presents anyone creating bibliometric data (the database producers) and undertaking bibliometric studies (the bibliometricians) with a special responsibility. The end users of the data are called upon to take the guidelines of both groups seriously to take account of the outcomes and relationships determined by bibliometric research over decades.

## Acknowledgements

We thank two anonymous reviewers for the recommendations to significantly improve the manuscript.

Table 1. Overview of the scientific performance of three researchers.

| Indicator | Person 1 | Person 2 | Person 3 |
|---|---|---|---|
| **Productivity** | | | |
|     Article | 143 | 54 | 43 |
|     Editorial | 1 | 1 | 4 |
|     Letter | 3 | 0 | 1 |
|     Meeting Abstract | 3 | 0 | 2 |
|     News Item | 0 | 2 | 0 |
|     Note | 12 | 0 | 1 |
|     Proceedings Paper | 26 | 17 | 40 |
|     Review | 2 | 2 | 4 |
| Total publications | 190 | 76 | 95 |
| Number of articles, notes, proceedings papers and reviews | 183 | 73 | 88 |
| Number of publications as first author[*] | 15 | 17 | 38 |
| Number of publications with no co-authors[*] | 0 | 5 | 12 |
| Year of first publication[*] | 1980 | 2001 | 1981 |
| Number of years between the first publication and 2011[*] | 32 | 11 | 31 |
| Number of publications per year (arithmetic average)[*] | 5.9 | 6.9 | 3.2 |
| **Impact** | | | |
| Total citations[**] | 15,192 | 3,796 | 7,828 |
| Number of citations per publication[**] (arithmetic average) | 83 | 52 | 89 |
| Proportion of self-citations in total citations[*] | 3.4% | 6% | 5.8% |
| h index[**] | 54 | 27 | 38 |
| m quotient[**] | 1.7 | 2.5 | 1.2 |
| Average percentile (median)[**] | 15.9 | 6.2 | 8.3 |
| $P_{top\ 10\%}$[**] | 70 | 31 | 48 |
| $PP_{top\ 10\%}$[**] | 39.3% | 52.5% | 57.8% |
| $P_{top\ 10\%}$ quotient[**] | 2.2 | 2.8 | 1.6 |

Remarks
[*] Based on publications of all document types
[**] Based on articles, letters, reviews, notes and proceedings papers

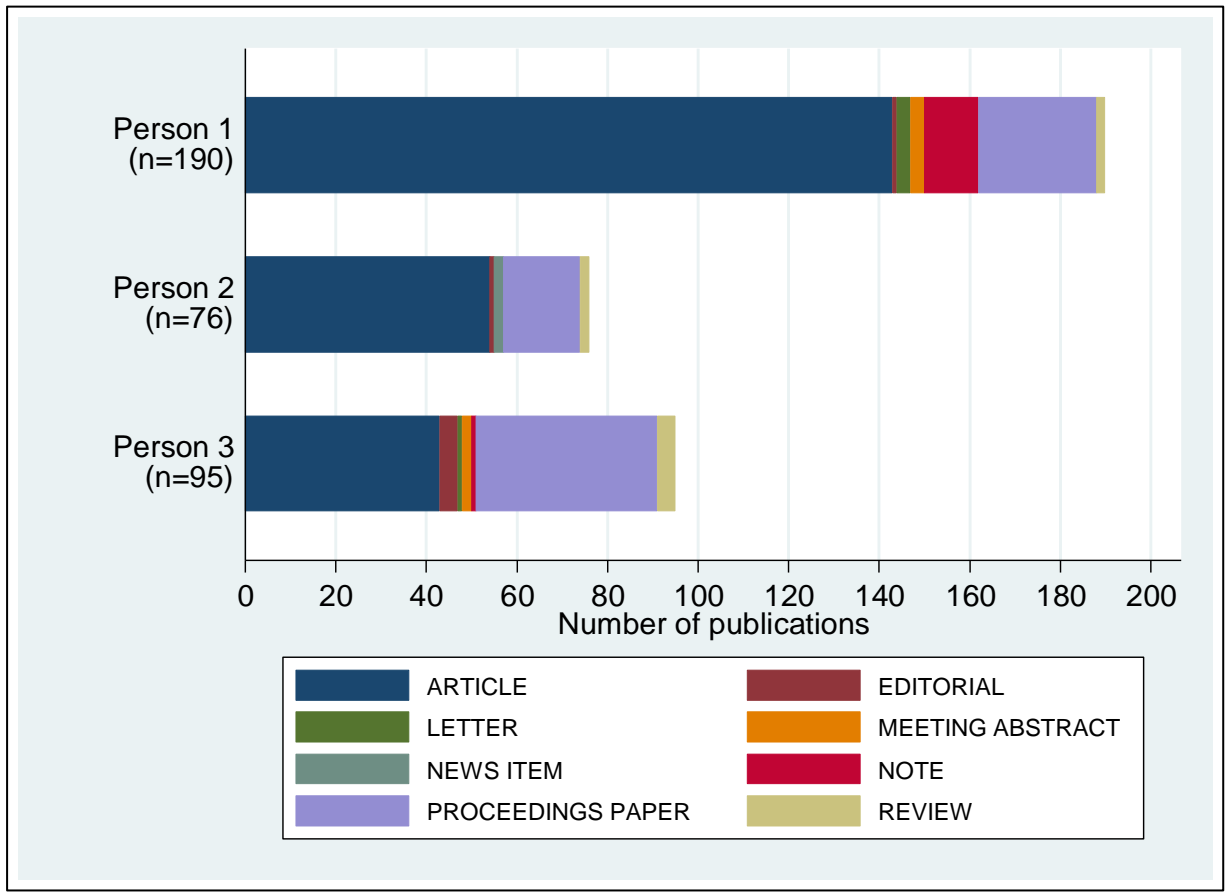

Figure 1. Number of publications with different document types by three researchers

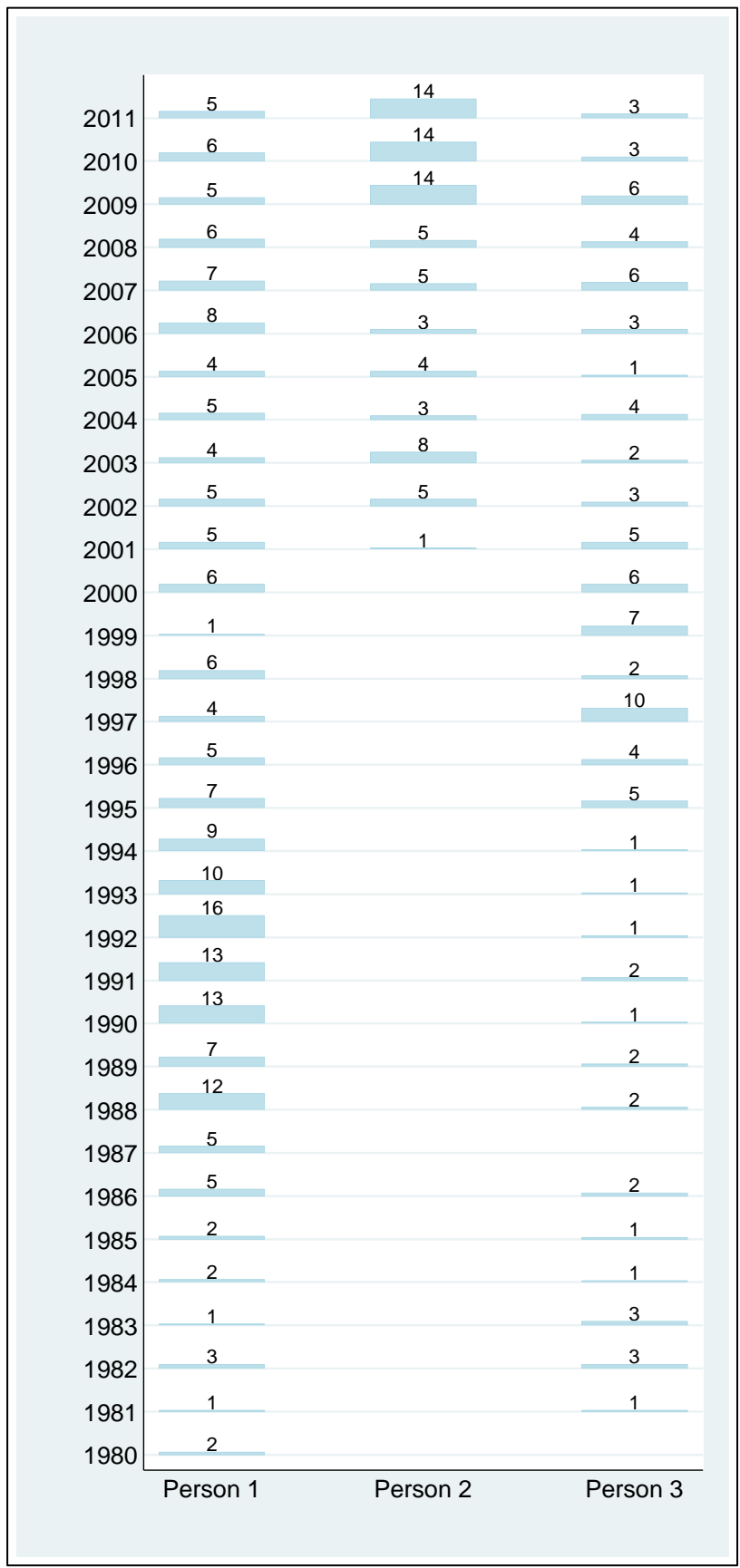

Figure 2. Number of publications by three researchers over the years

Table 2. Number of publications by three researchers in various journals. The Normalized Journal Position (NJP) based on the Journal Impact Factors from the Journal Citation Reports Science Edition 2011 is given for each journal. The journals are sorted in descending order by NJP.

| Person 1 | | | Person 2 | | | Person 3 | | |
| --- | --- | --- | --- | --- | --- | --- | --- | --- |
| Journal | Number | NJP | Journal | Number | NJP | Journal | Number | NJP |
| Journal 1 | 1 | 0.01 | Journal 1 | 3 | 0.01 | Journal 1 | 1 | 0.01 |
| Journal 2 | 3 | 0.05 | Journal 2 | 1 | 0.01 | Journal 2 | 1 | 0.03 |
| Journal 3 | 26 | 0.06 | Journal 3 | 9 | 0.03 | Journal 3 | 3 | 0.03 |
| Journal 4 | 1 | 0.07 | Journal 4 | 5 | 0.05 | Journal 4 | 1 | 0.05 |
| Journal 5 | 3 | 0.07 | Journal 5 | 1 | 0.05 | Journal 5 | 3 | 0.06 |
| Journal 6 | 1 | 0.07 | Journal 6 | 1 | 0.05 | Journal 6 | 1 | 0.08 |
| Journal 7 | 2 | 0.08 | Journal 7 | 3 | 0.06 | Journal 7 | 5 | 0.08 |
| Journal 8 | 1 | 0.08 | Journal 8 | 1 | 0.07 | Journal 8 | 3 | 0.09 |
| Journal 9 | 1 | 0.08 | Journal 9 | 1 | 0.07 | Journal 9 | 2 | 0.10 |
| Journal 10 | 1 | 0.10 | Journal 10 | 2 | 0.08 | Journal 10 | 1 | 0.14 |
| Journal 11 | 1 | 0.10 | Journal 11 | 1 | 0.09 | Journal 11 | 1 | 0.18 |
| Journal 12 | 1 | 0.16 | Journal 12 | 3 | 0.09 | Journal 12 | 4 | 0.20 |
| Journal 13 | 1 | 0.17 | Journal 13 | 1 | 0.09 | Journal 13 | 2 | 0.24 |
| Journal 14 | 72 | 0.19 | Journal 14 | 1 | 0.10 | Journal 14 | 3 | 0.24 |
| Journal 15 | 1 | 0.22 | Journal 15 | 2 | 0.11 | Journal 15 | 2 | 0.24 |
| Journal 16 | 1 | 0.22 | Journal 16 | 1 | 0.11 | Journal 16 | 3 | 0.26 |
| Journal 17 | 1 | 0.24 | Journal 17 | 2 | 0.13 | Journal 17 | 1 | 0.27 |
| Journal 18 | 6 | 0.26 | Journal 18 | 7 | 0.14 | Journal 18 | 1 | 0.30 |
| Journal 19 | 1 | 0.30 | Journal 19 | 1 | 0.17 | Journal 19 | 1 | 0.30 |
| Journal 20 | 1 | 0.30 | Journal 20 | 3 | 0.19 | Journal 20 | 2 | 0.31 |
| Journal 21 | 2 | 0.30 | Journal 21 | 1 | 0.23 | Journal 21 | 34 | 0.31 |
| Journal 22 | 1 | 0.34 | Journal 22 | 6 | 0.30 | Journal 22 | 2 | 0.38 |
| Journal 23 | 3 | 0.37 | Journal 23 | 2 | 0.34 | Journal 23 | 1 | 0.41 |
| Journal 24 | 1 | 0.41 | Journal 24 | 1 | 0.38 | Journal 24 | 1 | 0.43 |
| Journal 25 | 2 | 0.42 | Journal 25 | 1 | 0.47 | Journal 25 | 1 | 0.49 |
| Journal 26 | 1 | 0.42 | Journal 26 | 1 | 0.56 | Journal 26 | 2 | 0.62 |
| Journal 27 | 4 | 0.44 | Journal 27 | 1 | 0.59 | Journal 27 | 1 | 0.66 |
| Journal 28 | 4 | 0.45 | Journal 28 | 1 | 0.59 | Journal 28 | 1 | 0.88 |
| Journal 29 | 1 | 0.47 | **Total** | **63** | **0.19** | Journal 29 | 1 | 0.93 |
| Journal 30 | 1 | 0.49 | | | | **Total** | **85** | **0.29** |
| Journal 31 | 1 | 0.52 | | | | | | |
| Journal 32 | 1 | 0.59 | | | | | | |
| Journal 33 | 1 | 0.60 | | | | | | |
| Journal 34 | 2 | 0.63 | | | | | | |
| Journal 35 | 9 | 0.64 | | | | | | |
| Journal 36 | 2 | 0.64 | | | | | | |
| Journal 37 | 1 | 0.64 | | | | | | |
| Journal 38 | 1 | 0.70 | | | | | | |
| Journal 39 | 1 | 0.70 | | | | | | |
| Journal 40 | 1 | 0.77 | | | | | | |
| Journal 41 | 1 | 0.77 | | | | | | |
| Journal 42 | 1 | 0.80 | | | | | | |
| Journal 43 | 1 | 0.91 | | | | | | |
| **Total** | **169** | **0.36** | | | | | | |

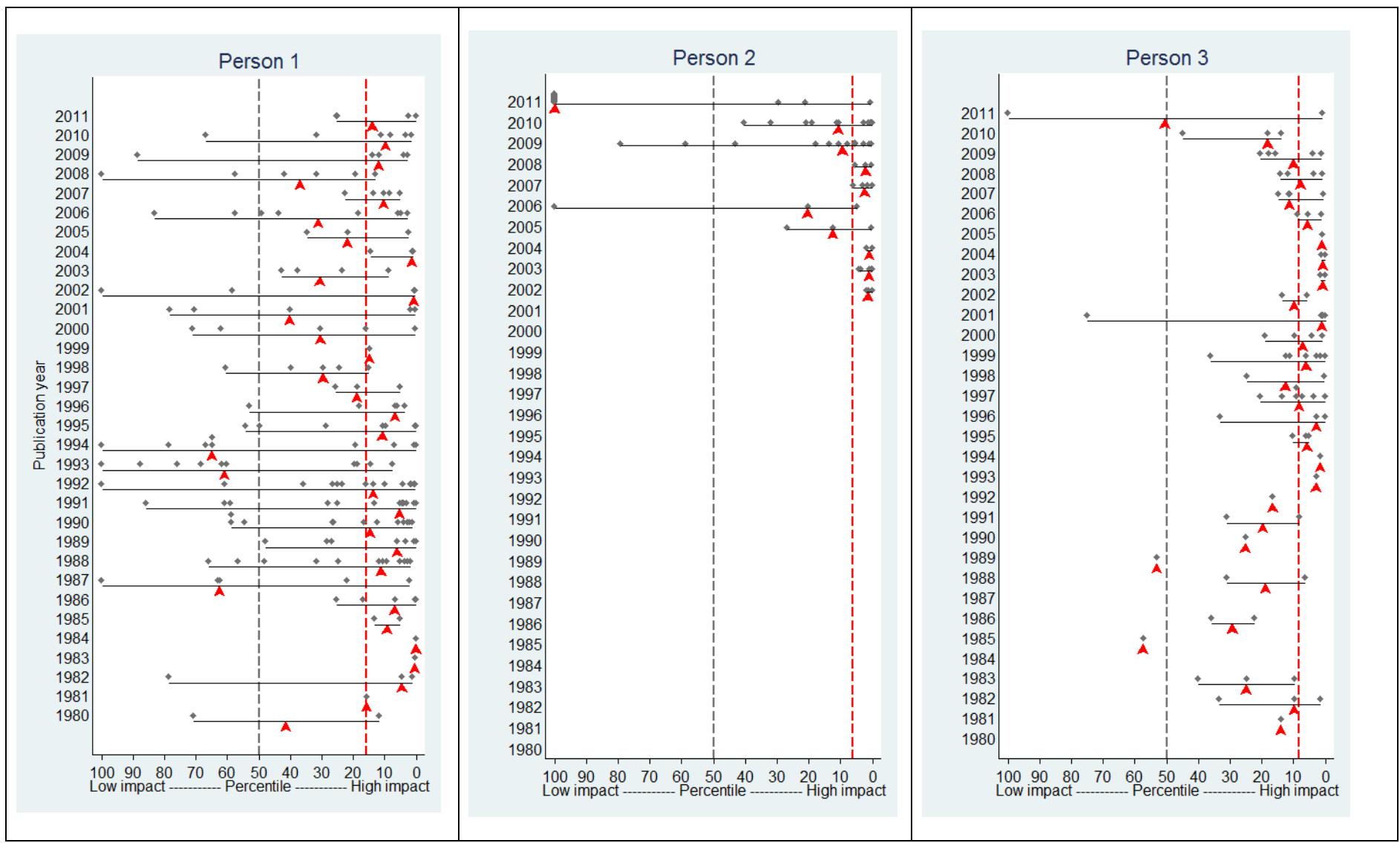

Figure 3. Distribution of percentiles for the publications by three researchers over the years